\documentclass[%
 reprint,
 amsmath,amssymb,
 aps,
]{revtex4-1}

\usepackage{graphicx}
\usepackage{dcolumn}
\usepackage{bm}

\begin{document}

\title{Topological weak-measurement-induced geometric phases revisited}%

\author{Graciana Puentes$^{1,2}$}
    \email[Correspondence email address: ]{gpuentes@df.uba.ar}
    \affiliation{1-Departamento de Fsica, Facultad de Ciencias Exactas y Naturales, Universidad de Buenos Aires, Ciudad Universitaria, 1428 Buenos Aires, Argentina,\\
    2-CONICET-Universidad de Buenos Aires, Instituto de Fsica de Buenos Aires (IFIBA), Ciudad Universitaria, 1428
Buenos Aires, Argentina. 
}

\date{\today} 

\begin{abstract}
We present an analytical and numerical study of a class of geometric phase induced by weak measurements. In particular, we analyze the dependence of the geometric phase on the winding $W$ of the polar angle $\varphi$, 
upon a sequence of $N$ weak measurements of increased magnitude ($c$), resulting in the appearance of a multiplicity of critical measurement-strength parameters where the geometric phase becomes stochastic. 
Adding to the novelty of our approach, we not only analyze the weak-measurement induced geometric phase by a full analytic derivation, valid in the quasicontinuous limit ($N \rightarrow \infty$), but also we analyze the induced geometric phase numerically, 
thus enabling us to unravel the finite-$N$ interplay of the geometric phase with the measurement strength parameter, and its stability to perturbations in the measurements protocol.
\end{abstract}

\maketitle

\section{Introduction}

When a quantum system undergoes adiabatic cycling, its state can become quantifiable based solely on its closed path in parameter space  \cite{1}. Conversely, a cyclic series of quantum observations can create a geometric phase. As reported in Gebhart \emph{et al.} \cite{2}, 
for closed trajectories the geometric phase becomes stochastic upon the application of a series of weak measurements, and a topological transition may occur in the mapping between the measurement sequence and the geometric phase, when the measurement strength is changed.

Despite the fact that overall quantum  phases cannot be fully determined, when the quantum system is driven slowly over a cycle returning to its starting condition, the accumulated phase becomes gauge invariant and can be measured. As originally pointed out by Sir. M. V. Berry 
\cite{3}, 
this is a geometric phase ($\mathcal{X}$) in that it is dependent on characteristics of the closed trajectory in parameter space, rather than on process dynamics. Geometric phases can be held responsible for a number of situations:  they can modify material properties in solids, such as conductivity in graphene \cite{Berrygraphene}, they can trigger the emergence of surface edge-states in topological insulators, whose surface electrons experience a geometric phase \cite{Berrytopoinsul},  they can modify the outcome of molecular chemical reactions  \cite{Berrychemestry}, and they can affect electronic properties of matter \cite{Berryelectronic}. Furthermore, understanding
 various physical phenomena  \cite{GeomPhase1, GeomPhase2, GeomPhase3}, defining fractional statistics anyonic quasiparticles \cite{Anyon1,Anyon2,Anyon3}, and identifying topological invariants for quantum Hall phases \cite{Hall1}, superconductors  \cite{TopoInsu1,TopoInsu2}, or quantitative characterizations of topological insulators via the Zak phase \cite{ZakDemler, ZakLonghi}, as well as underpinning  holonomic and topological signatures in photonic systems \cite{HolonomicWhite,PuentesCrystal,PuentesJOSAB,PuentesEntropy,PuentesQuantumRep}, are all made possible by geometric phases. 

\begin{figure} [h!]
\label{fig:1}
\includegraphics[width=1.0\linewidth]{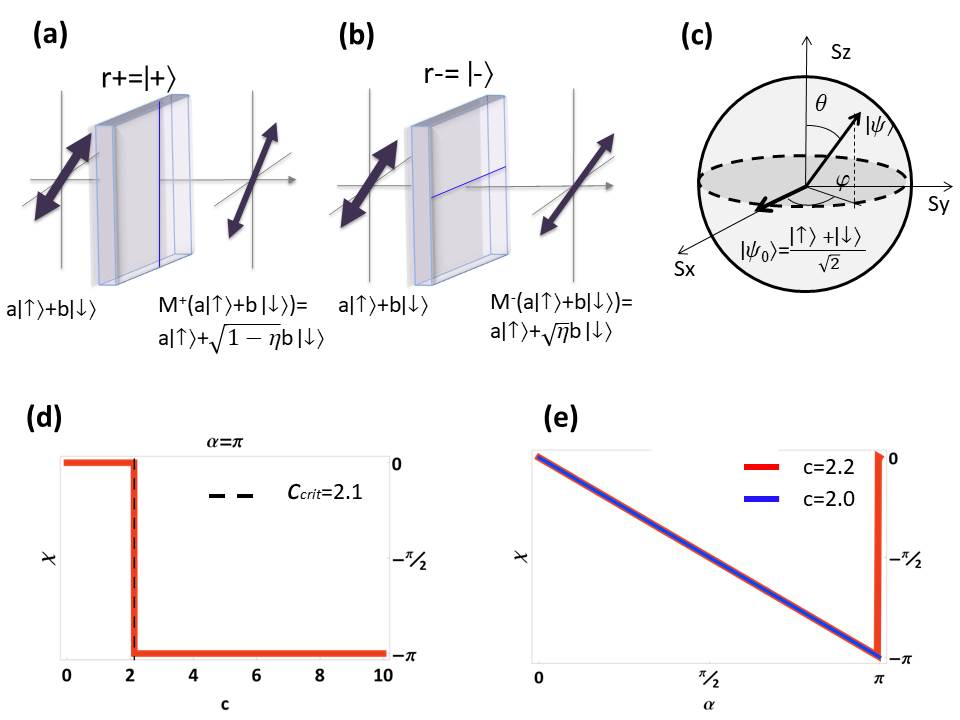}
\caption{Schematic representation of the action of the measurement protocol described by the proposed POVM set for a generic input state $a| \uparrow  \rangle + b| \downarrow \rangle$. Kraus operators $ M_{\eta}({\bf{n}},+/-)$ can be implemented by means 
of imperfect polarizers oriented along (a) vertical direction $r^{+}=|+\rangle$; (b) horizontal directions $r^{-}=|-\rangle$; (c) depicts the Bloch sphere for the system qubit $\{ | \uparrow  \rangle , | \downarrow \rangle \}$ setting 
the initial state $| \psi_0 \rangle=\frac{|\uparrow \rangle + |\downarrow \rangle}{\sqrt{2}}$, for initial parameters $( \theta_0=\pi /2,\varphi_0=0)$; (d) red curve depicts the discrete jump in the geometric phase $\mathcal{X}$ from 0 to $-\pi$ for a 
critical measurement strength $c_{\mathrm{crit}}=2.1$ (dashed line indicating the onset of topological phase transition; (e) depicts the topological phase transition in $\mathcal{X}$ for $\alpha=\pi$, corresponding to a single winding of $\varphi$ ($W=1$).  Special attention was taken to ensure the continuity of $\mathcal{X}$ at the point of the stochastic phase transition in order to distinguish transitions from random phase jumps.}
\end{figure}

\begin{figure} [t!]
\label{fig:1}
\hspace{1cm}
\includegraphics[width=1.0\linewidth]{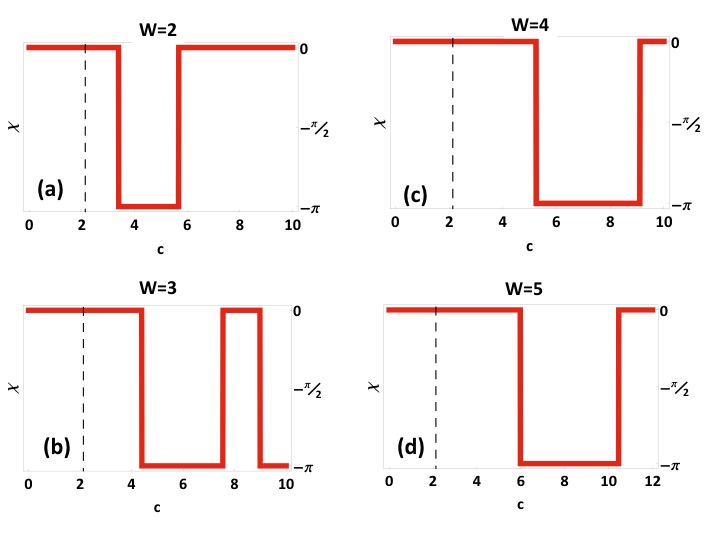}
\vspace{-0.4cm} \caption{Stochastic transition in the geometric phase $\mathcal{X}$ vs. angle parameter $\alpha$ for increased winding number $W$: (a) $(\alpha= 2 \pi, W=2)$ with $c_{\mathrm{crit}}=3.4$ and $5.7$, (b) $(\alpha= 3 \pi, W=3)$ with  $c_{\mathrm{crit}}=4.4$ and $7.6$, (c) ($\alpha= 4 \pi, W=4$) with $c_{\mathrm{crit}}=5.2$ and $9.1$ and  (d)  $(\alpha= 5\pi, W=5)$ with $c_{\mathrm{crit}}=6.0$ and $10.5$. We note the actual critical measurement-strength values are completely unpredictable, thus the transitions at different winding numbers $W$ are not topologically equivalent. }
\end{figure}

 A class of geometric phase, resulting from the outcome of a series of intense (projective) measurements that operate on the system and produce certain measurement read-outs, is the Pancharatnam phase \cite{Pancharatnam}. Optical investigations that monitor the Pancharatnam phase 
 caused by polarizer sequences have readily been reported \cite{BerryStack}. Notwithstanding the fact that incoherent measurement processes are generally  involved, such a phase may be reliably detected. A general series of measurements is by its very nature stochastic. 
 Thus, based on the sequences of measurement readouts linked to the relevant probabilities, one may expect a distribution of measurement-induced geometric phases. The induced evolution is entirely predictable for a quasi-continuous series of strong measurements ($N \rightarrow \infty$) because of the dynamical quantum Zeno effect \cite{Zeno}.\\

In this paper, we present an analytical and numerical study of a novel type of geometric phase induced by weak measurements. In contrast to PNAS 2020 \cite{2}, the originality of our approach consists of considering the dependence of the geometric phase on the winding $W$ of the
 polar angle $\varphi$, which quantifies the number of full $2\pi$-turns the trajectory makes until it closes its path, returning to its original state. This is accomplished by fixing the azymhutmal angle at $\theta=\pi/2$ and considering the initial state 
 $|\psi_0 \rangle = \frac{| \uparrow  \rangle + | \downarrow \rangle}{[\sqrt{2}}$, and a sequence of weak measurements in the angle $\varphi$ of increasing magnitude $\varphi_{k}= \epsilon k$, with $k=0,..,N$ the measurement index. 
 We ensure the trajectory induced by the sequence of weak measurements is closed, and the geometric phase well defined, by parameterizing the rotation parameter as $\alpha=2 \epsilon/N$, 
 where $N$ is the number of measurements. With this parameterization, the winding number is simply defined as $W=\alpha/\pi$, resulting in $W=1,...,M$ with $M$ the total number of windings. We find that different winding numbers $W$ can give rise to unpredictable critical measurement-strength parameters, 
 where the geometric phase becomes stochastic and the system undergoes a topological phase transition, thus confirming that the transitions for different $W$ are not topologically equivalent.  
 Furthermore, adding to the novelty of our work, we not only analyze the weak-measurement induced geometric phase by a full analytic derivation  based on the exponential approximation presented in PNAS 2020 \cite{2}, valid for $N \rightarrow \infty$, 
 but also we analyze the induced geometric phase numerically, thus enabling us to understand the finite-$N$ interplay of the geometric phase on the measurement strength parameter $c$, and its stability to fluctuations in the measurement protocol.  \\

The paper is organized as follows: In Section II we outline the measurement protocol. Next, in Section III, we present our analytic results. In particular, we characterize the phase transitions for different windings $W$. Next, in Section IV, we present our numerical results. Namely, we characterize the interplay of the geometric phase with the measurement strength $c$ for finite $N$, and we analyze the impact of phase noise on the geometric phase in different parameter regions of the $N-c$ landscape. Finally, in Section V, we outline the conclusions and future perspectives.

\section{Measurement Protocol}

\begin{figure*}
\label{fig:1}
\includegraphics[width=1\linewidth]{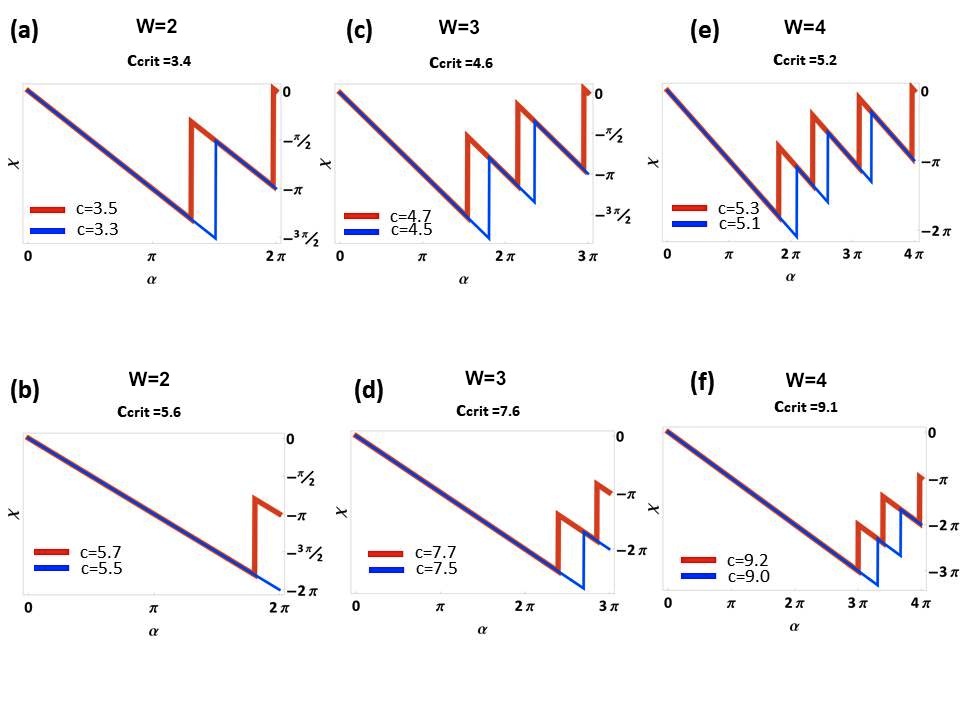}
\vspace{-0.4cm} \caption{Topological phase transition for increased winding numbers $W=2,3,4$, for $c >c_{\mathrm{crit}}$ ($c < c_{\mathrm{crit}}$) red(blue) curves: (a) and (b) depicts curves for $W=2$, and phase transition from 0 to $|\pi|$ and $|\pi|$ to 0, respectively; (c) and (d) depicts curves for $W=3$, and phase transition from 0 to $|\pi|$ and $|\pi|$ to 0, respectively; (e) and (f) depicts curves for $W=4$, and phase transition from 0 to $|\pi|$ and $|\pi|$ to 0, respectively. The different topological nature of each phase transition is signalled by the increased numbers of oscillations for increased $W$. This can be explained by noting that increasing the winding number $W=k \pi$ corresponds to closing the trajectory after $k$-windings of the polar phase  $\phi$. Therefore, it is expected to observe a larger number of oscillations when for increased $W$. (a), (c) and (e) display $k$ oscillations for $W=k \pi$, while (b), (d) and (f) present $k-1$ oscillations for $W=k \pi$, since there is no phase transition from $|\pi|$ to 0 for a single winding $W=1$.  Special attention was taken to ensure the continuity of the geometric phase at the point of the stochastic transition in order to distinguish transitions from random phase jumps.  }
\end{figure*}

The measurement sequence required to accumulate the intended geometric phase $\mathcal{X}$ can be mathematically described by a complete set of POVMs (Positive Operator Valued Measures),
 implemented via the Krauss operators $\mathcal{M}_{k}^{(r_{k})}=M_{\eta_{k}}({\bf{n_{k}}},r_{k})$, $|\psi \rangle \rightarrow \mathcal{M}_{k}^{(r_{k})} |\psi \rangle$, as described in \cite{2}. 
 Such POVM can be implemented by introducing a detector consisting of a second qubit whose Hilbert space is spanned by the set $r=\{ |+ \rangle, |- \rangle\}$. We consider the generic initial state of the system of the form $|\psi_0 \rangle = a |\uparrow \rangle +b|\downarrow \rangle$, and assume the detector is in the initial state $|+ \rangle$ and that the initial state of the system plus detector is separable, of the form $|\psi_{\mathrm{sep}}\rangle = |\psi_0 \rangle \otimes |+ \rangle$. The measurement coupling $\lambda(t)$ is then switched on for a finite time $t \in [0,T]$, to obtain the entangled state:

\begin{equation}
|\psi_{\mathrm{ent}}\rangle = M_{\eta}({\bf{n}},+) |\psi_0 \rangle +  M_{\eta}({\bf{n}},-) |\psi_0 \rangle,
\end{equation}

here the measurement strength is $\eta= \sin^2(g)$, with $g=\int_0^{T} \lambda(t)dt$. The POVM set is defined by the Kraus operators:

\begin{equation}
 M_{\eta}(\hat{z},+)= \begin{pmatrix}
 1 &  0\\
 0 &  \sqrt{1-\eta}
  \end{pmatrix} \hspace{0.5cm}   M_{\eta}(\hat{z},-)= \begin{pmatrix}
 1 &  0\\
 0 &  \sqrt{\eta}
  \end{pmatrix}
  ,
  \end{equation}
  
 corresponding to a measurement orientation along the $z$-axis ${\bf n}=\hat{z}$. Kraus operators along a generic orientation ${\bf n}$ can be obtained via the change of basis $ M_{\eta}({\bf n},r)=R^{-1}({\bf n}) M_{\eta}(\hat{z},r)R({\bf n}) $, where the unitary matrix $R({\bf n})$ is given by:
 
 \begin{equation}
 R({\bf n}) =
 \begin{pmatrix}
 \cos \theta/2 & e^{-i \varphi}\sin \theta/2\\
 \sin \theta/2 & e^{-i \varphi}\cos \theta/2
  \end{pmatrix}
  ,
 \end{equation}
 
 representing a rotation of the measurement orientation along the direction ${\bf n}$. This map implements a null-type weak measurement as proposed in PNAS2020 \cite{2}. An imperfect polarizer oriented along the directions $r^{+,-}$ can implement such null-type weak-measurement protocol. This is depicted in Figure 1. Fig. 1 (a) describes the action of the POVM $M_{\eta}(\hat{z},+)$ on a generic input state $|\psi_0 \rangle =a | \uparrow  \rangle + b | \downarrow \rangle$  while Fig. 1 (b) describes the action of the POVM $M_{\eta}(\hat{z},-)$ 
 on the same generic input state. The Bloch sphere for the system qubit $s=\{ |\uparrow \rangle, |\downarrow \rangle\}$ is schematized in Fig. 1 (c). 
  
 \bigskip
  
 \section{Analytic Results}
 
\begin{figure} [t!]
\label{fig:1}
\hspace{0.8cm}
\includegraphics[width=1\linewidth]{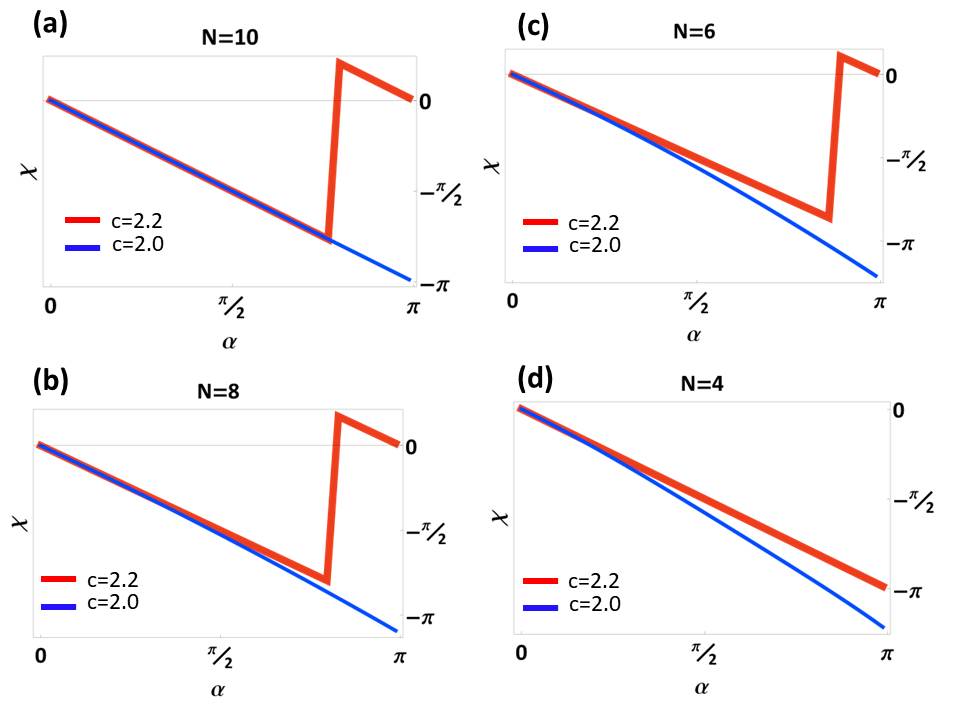}
\vspace{-0.4cm} \caption{Plots of the geometric phase $\mathcal{X}$ vs $\alpha$ for decreasing values of $N$, for a measurement strength parameter $c>c_{\mathrm{crit}}$ (red curves) and  $c<c_{\mathrm{crit}}$ (blue curves), 
where $c_{\mathrm{crit}} \approx 2.1$ for $W=1$.  (a), (b), (c) and (d) correspond to $N=10,8,6,4$, respectively. The deviation of the numerical result from the analytic prediction is readily apparent for $N=8$. For $N=4$ the stochasticity in  $\mathcal{X}$ is fully vanished.  }
\end{figure}

\begin{figure*} [t!]
\label{fig:1}
\hspace{0.8cm}
\includegraphics[width=1\linewidth]{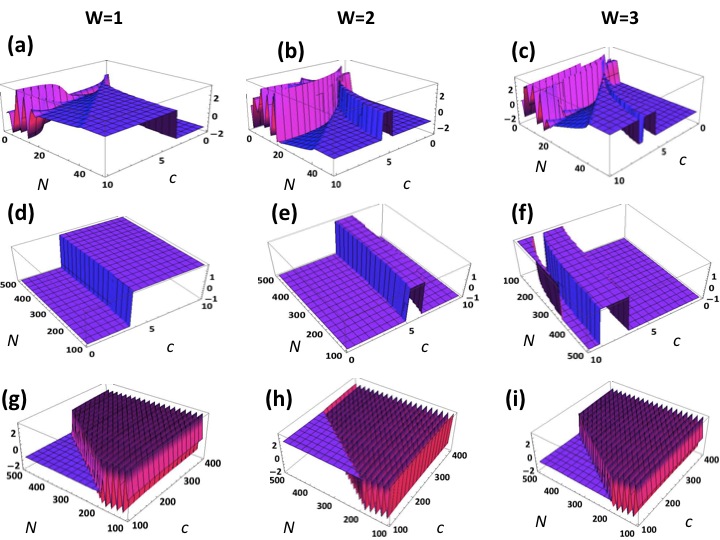}
\vspace{-0.4cm} \caption{3D plots of the geometric phase $\mathcal{X}$ for different regions of the $N-c$ landscape. (a), (b) and (c) correspond to parameter values of  $0 < N < 50$ and $0 < c < 10$, for $W= 1,2,3$ respectively. The magenta colouring  of the 3D density plots indicate regions where the stability of the geometric phase is dramatically reduced. (d), (e) and (f) present numerical simulations of  $\mathcal{X}$ considering $100< N < 500$  and $0 < c < 10$, for $W= 1,2,3$ respectively. (g), (h), (i), for $100 < N < 500$ and $100 < c < 400$, for $W=1,2,3$, respectively. It is observed that fluctuations in $\mathcal{X}$ arise  for $c >300$, for any value of $N$. Moreover, in the stable region characterized by $N>>c$, there is no apparent critical measurement-strength parameter where $\mathcal{X}$ makes a discrete jump, within the resolution of the numerical simulations.  }
\end{figure*}

 An analytic expression for the geometric phase $\mathcal{X}$ can be derived in the quasicontinuous measurement limit $N \rightarrow \infty$. $\mathcal{X}$ is extracted from the quasicontinuous trajectory postselecting all outcomes $r_{k} = |+\rangle$ as described in PNAS2020 \cite{2}. 
 This result is obtained by setting the initial state $| \psi_0 \rangle= R^{-1}({\bf n_0}) | \uparrow \rangle$. In our case, we consider the initial state to be eigenstate of $S_{x}$, of the form $|\psi_0 \rangle = \frac{| \uparrow  \rangle + | \downarrow \rangle}{[\sqrt{2}}$, 
 thus selecting the initial rotation along $\hat{x}$, this is equivalent to setting the initial parameters $(\theta_0=\pi/2, \varphi_0=0)$, as depicted in Fig. 1 (c). We then sequentially rotate the measurement apparatus, in order to increment the angle $\varphi$ 
 by a fixed amount $\epsilon=2 \pi/N$. This parameterization ensures that the trajectory is closed and the geometric phase well defined. By selecting the readouts $r_{k}= |+ \rangle$, setting $\eta=4c/N$, 
 the measurement orientations $(\theta_{k} , \varphi_{k} ) = (\pi/2, 2\pi k/N)$, and using the explicit form of Kraus operators in Eq. 2, one obtains an expression for the geometric phase $\mathcal{X}$  of the form \cite{2}:
 
 \begin{equation}
 \langle \psi_0 | \mathcal{M}^{(+)}_{N-1}...\mathcal{M}^{(+)}_{1}  | \psi_0 \rangle=   \langle \uparrow| \delta R (M_{4c/N}(\hat{z},+)\delta R)^{N-1}| \uparrow \rangle,
 \end{equation}

where $\delta R$ is given by the $k$-independent matrix:

 \begin{equation}
 \delta R =
 \begin{pmatrix}
 1/2(1+e^{-i \epsilon}) &  1/2(1-e^{-i \epsilon})\\
  1/2(1-e^{-i \epsilon}) &  1/2(1+e^{-i \epsilon})
  \end{pmatrix}
.
 \end{equation}
 
 The quasicontinuous limit ($N \rightarrow \infty$) is obtained by using the exponential approximation $(1+ \Gamma /N)^N \approx e^{\Gamma}$, valid for $N \rightarrow \infty$. This approximation is explicitly invoked by expressing the Taylor series of  $M_{4c/N}(\hat{z},+) \delta R$, and rewriting  $M_{4c/N}(\hat{z},+) \delta R=V (I + \Gamma/N)V^{-1}$, where $I$ is the $2 \times 2$ identity matrix and  $V$ is the unitary matrix for the change of basis, obtained by diagonalization of $ M_{4c/N}(\hat{z},+) \delta R$. In this scenario, the geometric phase can be extracted from the upper-most diagonal element of the matrix product $V e^{\Gamma} V^{-1}$, resulting in:
 
 \begin{equation}
 \mathcal{X}= e^{-i \alpha -c}[\cosh(\tau) + c \sinh(\tau)/\tau),
 \end{equation}
 
 where $\tau=\sqrt{c^2 - \alpha^2}$ and $\alpha= \epsilon N/2$. We recover the analytic result reported in  \cite{2}, by setting $z=c$, and $\alpha=\pi$, as expected for $\theta=\pi/2$. \\
 
We note that our case is complementary to the one analyzed in PNAS2020 \cite{2}, where the authors analyze the dependence of the geometric phase on the azymhutmal angle $\theta$ for a single winding ($ \alpha=\pi$) of the polar angle $\varphi$. Here we report the dependence of the geometric phase on the winding of the polar angle $\varphi$, obtained by increasing the  winding number $W=\alpha/\pi$, while fixing $\theta=\pi/2$. We stress that the critical measurement-strength parameter $c_{\mathrm{crit}}$ obtained for different winding numbers $W=1,2, 3,...,M$ (with $M \in N$) are completely random. Therefore, we argue the phase transitions observed for different windings $W$ are not topologically equivalent. \\
 
The onset of the topological phase transition in the geometric phase for a critical measurement strength can be understood by considering two limiting cases: For the case of a series of strong projective measurements ($c \rightarrow \infty$), resulting in the well-known Pantcharatnam geometric phase \cite{Pancharatnam}, 
the polarization is expected to rotate along the equator by an amount $\epsilon=2 \pi/N$ with each projective measurement, until it returns to its original state, accumulating a geometric phase equal to 1 (modulo $2 \pi$), 
while for a series with infinitely weak measurement strength ($c \rightarrow 0$) the polarization should not be affected by the measurement process. Thus, there exists a critical measurement strength $c_{\mathrm{crit}}$ where $\mathcal{X}$ makes a discrete jump from 0 to 1, 
signalling the onset of a topological phase transition, here $\pi$ corresponds to a full winding of the geometric phase. A plot of the geometric phase $\mathcal{X}$, given by the analytic expression in Eq.  6, is given in Fig. 1 (d) and (e). Figure 1 (d) 
depicts the discrete jump of $\mathcal{X}$ from 0 to $-\pi$ for a critical measurement strength $c_{\mathrm{crit}}=2.1$ indicating the onset of topological phase transition, while Fig. 1 (e) depicts the geometric phase vs. $\alpha$, signalling a topological phase transition 
for $\alpha=\pi$, corresponding to a full winding of the polar phase ($W=1$). Special attention was taken to ensure that the geometric phase is mathematically continuous in order to distinguish the onset of a topological transition, from a random $\pi$-jump in the geometric phase.

\subsection{Winding Number ($W=\alpha / \pi$)}

Next, we analyzed the existence of additional critical measurement-strength values, signalling multiple topological phase transition characterized by discrete jumps in the geometric phase between 0 and $| \pi|$. This was analyzed by enabling the parameter $\alpha$ to take  multiple values  $k \pi$ ($k=1,...,M$ with $M \in N$), corresponding to a full winding of the phase $\varphi$ quantified by the winding number $W= \alpha / \pi=1,...,M$ and a wrapping of the phase at multiples of $2 \pi$, revealing the existence of additional critical measurement-strength parameters for different values of $W$. This is shown in Figure 2, displaying additional critical-measurement parameter values where 
$\mathcal{X}$ jumps from $0 \rightarrow - \pi$ and from $- \pi \rightarrow  0$, respectively.
We note that the critical values $c_{\mathrm{crit}}$ are fully random, therefore the phase transitions observed for different windings $W$ are not topologically equivalent. 
Figure 2 corresponds to: (a) $(\alpha= 2 \pi, W=2)$ with $c_{\mathrm{crit}}=3.4$ and $5.7$, (b) $(\alpha= 3 \pi, W=3)$ with  $c_{\mathrm{crit}}=4.4$ and $7.6$, (c) ($\alpha= 4 \pi, W=4$) with $c_{\mathrm{crit}}=5.2$ and $9.1$ and  (d)  $(\alpha= 5\pi, W=5)$ 
with $c_{\mathrm{crit}}=6.0$ and $10.5$. We note the actual critical measurement-strength values are completely unpredictable, thus we argue the topological transitions indicated by different winding numbers $W$ are not equivalent. \\

Further confirmation of the onset of different topological phase transition was obtained by plotting the geometric phase vs. the angle parameter $\alpha$ for increased winding number $W$. 
This is presented in Figure 3, for trajectories with $c<c_{\mathrm{crit}}$ (red curves) and for trajectories with $c>c_{\mathrm{crit}}$ (blue curves).  Figure 3 (a) and (b) depicts curves for $W=2$,
 and phase transition from 0 to $|\pi|$ and $|\pi|$ to  $0$ (mod $2\pi$), respectively; (c) and (d) depicts curves for $W=3$, and phase transition from 0 to $|\pi|$ and $|\pi|$ to 0  (mod $2\pi$), respectively; (e) and (f) depicts curves for $W=2$, 
 and phase transition from 0 to $|\pi|$ and $|\pi|$ to  $0$  (mod $2\pi$), respectively. The different topological nature of each phase transition is signalled by the increased numbers of oscillations for increased $W$. 
 This can be explained by noting that increasing the winding number $W=k \pi$ corresponds to closing the trajectory after $k$-windings of $2 \pi$ of the polar angle  $\varphi$. Therefore, it is expected to observe a larger number of oscillations for increased 
 windings $W$, as displayed in Fig. 3. Figure 3 (a), (c) and (e) display $k$ oscillations for $W=k \pi$, while Figs. 3 (b), (d) and (f) present $k-1$ oscillations for $W=k \pi$, since there is no phase transition from $|\pi|$ to 0 (modulo  2$\pi$) for a single winding $W=1$.

\begin{figure*} [t!]
\label{fig:1}
\hspace{0.8cm}
\includegraphics[width=1\linewidth]{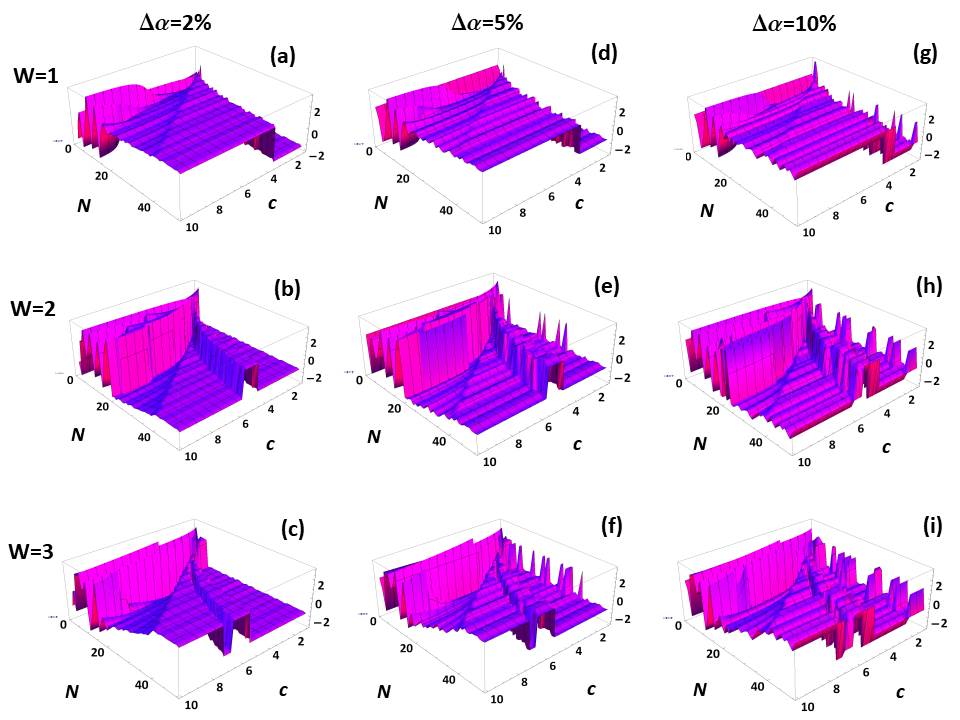}
\vspace{-0.4cm} \caption{Numerical simulations of uncorrelated phase noise in the parameter region $0<N<50$ and $0<c<10$ of the $N-c$ landscape.  (a), (b), (c) correspond to phase noise with a spread $\Delta \alpha = 2\%$ for $W=1,2,3$, respectively; (d), (e), (f) correspond to phase noise with a spread $\Delta \alpha = 5\%$ for $W=1,2,3$, respectively; (g), (h), (i) correspond to phase noise with a spread $\Delta \alpha = 10\%$ for $W=1,2,3$, respectively. For $\Delta \alpha > 2 \%$ fluctuations in the phase render the measurement protocol significantly unstable.   
}
\end{figure*}

\begin{figure*} [t!]
\label{fig:1}
\hspace{0.8cm}
\includegraphics[width=1\linewidth]{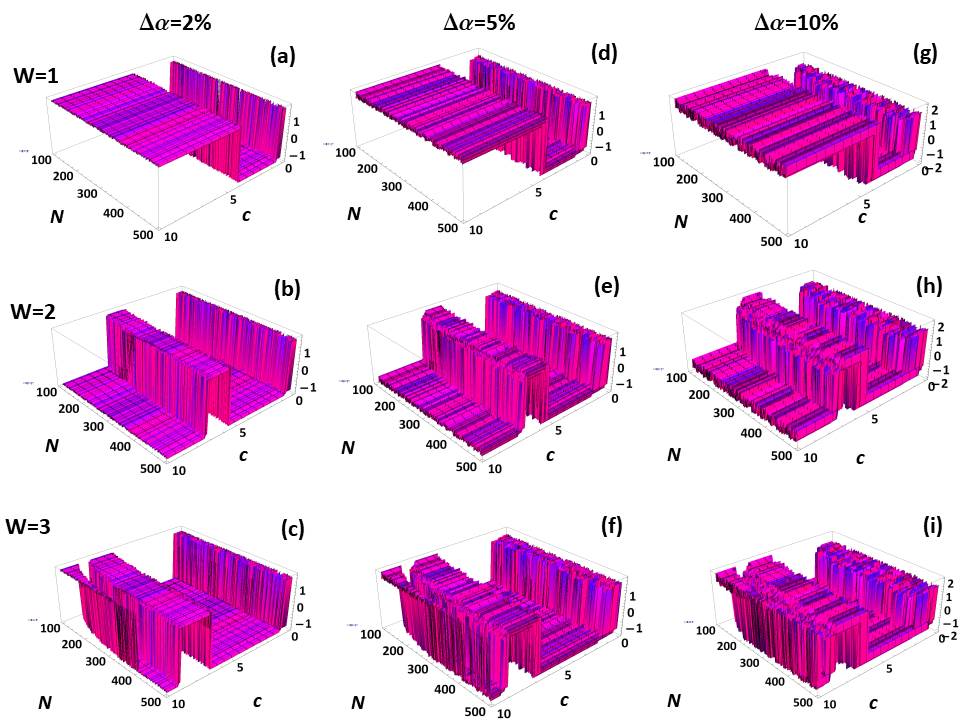}
\vspace{-0.4cm} \caption{Numerical simulations of uncorrelated phase noise in the parameter region $100<N<500$ and $0<c<10$ of the $N-c$ landscape. (a), (b), (c) correspond to phase noise with a spread $\Delta \alpha = 2\%$ for $W=1,2,3$, respectively; (d), (e), (f) correspond to phase noise with a spread $\Delta \alpha = 5\%$ for $W=1,2,3$, respectively; (g), (h), (i) correspond to phase noise with a spread $\Delta \alpha = 10\%$ for $W=1,2,3$, respectively. This region of the $N-c$ landscape is significantly robust to uncorrelated phase noise and the quantization of the phase is typically  preserved.}
\end{figure*}

\begin{figure*} [t!]
\label{fig:1}
\hspace{0.8cm}
\includegraphics[width=1\linewidth]{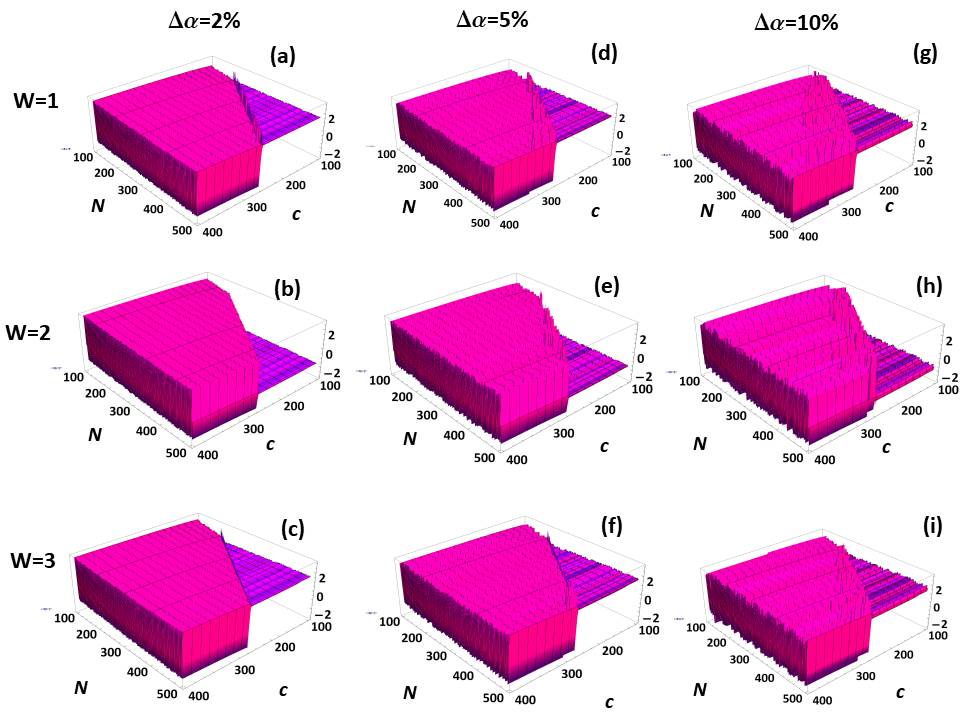}
\vspace{-0.4cm} \caption{Numerical simulations of uncorrelated phase noise in the parameter region $100<N<500$ and $100<c<400$ of the $N-c$ landscape. (a), (b), (c) correspond to phase noise with a spread $\Delta \alpha = 2\%$ for $W=1,2,3$, respectively; (d), (e), (f) correspond to phase noise with a spread $\Delta \alpha = 5\%$ for $W=1,2,3$, respectively; (g), (h), (i) correspond to phase noise with a spread $\Delta \alpha = 10\%$ for $W=1,2,3$, respectively. In this limit, the only region of the $N-c$ landscape robust to phase noise if for $N>>c$ and $c<300$, even though there are no apparent phase transitions within the resolution of the numerical simulations. }
\end{figure*}

\section{Numerical Results}

\subsection{Quantization of the geometric phase}

The topological character of the phase transition exhibited at the critical measurement-strength parameter ($c_{\mathrm{crit}}$), where the geometric phase  ($\mathcal{X}$) becomes stochastic, is clearly revealed by the quantization of $\mathcal{X}$ 
between 0 and $|\pi|$ (modulo $2\pi$). This quantization can be understood as a signature of the so-called ``topological protection'' of the phase, which prevents fluctuations in the discrete values of the phase acquired by the system 
 under weak perturbations in the protocol. This quantization of the phase, confirming the topological nature of the transition,  is clearly valid in the quasicontinuous limit ($N \rightarrow \infty$), where the analytic result holds. \\
 
In this Section, we analyze the robustness of the quantization of $\mathcal{X}$, with respect to fluctuations in the total number of measurements $N$, meaning that we derive a numerical expression for the geometric phase for  finite-$N$. Evidently, this analysis cannot be derived from the analytical result reported in Eq. (6), which is only valid for $N \rightarrow \infty$ and $N>> \Gamma$, for the exponential approximation $(1+ \Gamma/N)^N \approx e^{\Gamma}$ to hold. 
In order to obtain a numerical value for $\mathcal{X}$ and analyze its robustness to different type of noise and perturbations, we follow Eq. (4) to obtain a numerical expression for the $N$-matrix product  
$\delta R (M_{4c/N}(\hat{z},+)\delta R)^{N-1}$. Next, we plot the left upper-most matrix element for different values of $N$ and $c$. For $N$ sufficiently large ($N>500$) and the measurement-strength parameter sufficiently small ($c<<N$), the analytic result and numerical results are indistinguishable. 
We are interested in understanding at which point the numerical result departs from the analytic prediction.\\

Plots of the geometric phase $\mathcal{X}$ vs $\alpha$ for decreasing values of $N$ are presented in Fig. 4, for a measurement strength parameter $c>c_{\mathrm{crit}}$ (red curves) and  $c<c_{\mathrm{crit}}$ (blue curves), where $c_{\mathrm{crit}} \approx 2.1$ for $W=1$. For simplicity, we 
consider a single winding of the polar angle $W=1$, although the same framework can be applied to $W>1$. Figures 4 (a), (b), (c) and (d) correspond to $N=10,8,6,4$, respectively. We note that the deviation of the numerical result from the analytic prediction is readily apparent for $N=8$, 
where the quantization of $\mathcal{X}$ between 0 and $\pi$ vanishes, signalling that the topological nature of the phase transition vanishes. Furthermore, for $N<5$ the stochastic response of the geometric phase at the critical measurement strength is completely washed out. 
Our numerical findings confirms that the quantization and stochasticity of ($\mathcal{X}$) not only depends on the measurement strength $c$, but also on the total number of measurements $N$ and its interplay with the $c$-parameter, with fully stochastic character only for $N>4$, when considering $c_{\mathrm{crit}}=2.1$ ($W=1$).

\subsection{$N-c$ landscape} 

In order to analyze the interplay between the quantization of the geometric phase $\mathcal{X}$, for different values of total number of measurements $N$ and strength parameter $c$, we numerically calculated $\mathcal{X}$ in different relevant regions of the $N-c$ landscape. This is displayed in the 3D density plot depicted in Fig. 5. Figure 5 (a), (b) and (c) correspond to parameter values of  $0 < N < 50$ and $0 < c < 10$, for $W= 1,2,3$ respectively. Clearly, it is observed that the stability of the geometric phase is dramatically  reduced
 for $N<20$ (magenta regions), meaning that fluctuations in the geometric phase are readily apparent for $N$ of the same order as $c$ ($ N\approx c$). This numerical finding also sets a limit in the actual values of measurement strength $c$  that can be accepted in the quasicontinuous limit. Figures 5 (d), (e) and (f) present numerical simulations of  $\mathcal{X}$ considering $100< N < 500$  and $0 < c < 10$, for $W= 1,2,3$ respectively. We find that for $N >100$, corresponding to $N>> c$, the fluctuations in the geometric phase vanish,  meaning that in this region of the $N-c$ landscape the analytic result is fully valid, as reported in \cite{2}. Fluctuations are expected to resurge as $c$ approaches values of the same order of $N$  ($ N\approx c$) \cite{19}, or for values larger than $N$ ($N < c$). This is analyzed in detail in Fig. 5 (g), (h), (i), for $100 < N < 500$ and $100 < c < 400$, for $W=1,2,3$, respectively. We note that there is no clear critical measurement-strength parameter in this region where $\mathcal{X}$ makes a discrete jump, as a result of the limited resolution in $c$ required in order to survey such a large portion of the $N-c$ landscape. Nevertheless, it is observed that fluctuations in $\mathcal{X}$ arise  for $c > 300$, for any value of $N$, meaning that the region of validity of the analytical result can only be considered if $N>>c$. In other words, $N>c$ is not acceptable for the analytic result to hold. Evidently, there is no value of $N$ where the analytic result holds for $c > 300$. This upper limit is relevant when analyzing the strong-measurement limit  $c \rightarrow \infty$, as considered in Snizkho \emph{et al.} \cite{19}, and for feasible experimental realizations of the scheme \cite{21,22}.  

\subsection{Phase Noise} 

In order to characterize the robustness of the measurement protocol to experimental errors arising as a result of phase fluctuations due to limited temporal stability of the system, which can result for example as a consequence of temperature, spatial, frequency, or polarizations drifts, among other realistic sources of errors. We analyze the effect of uncorrelated phase noise on the geometric phase $\mathcal{X}$ in the $N-c$ landscape, although other forms of correlated noise could be considered, for instance noise linearly or polynomially correlated with the total number of measurements $N$, and the measurement strength $c$. We model the phase noise $\Delta \alpha$ with a normal distribution centred around $\alpha_0=\pi$, with varying spreads $\Delta \alpha=2 \%, 5\%, 10 \%$, considering different winding numbers for the polar angle $W=1,2,3$. Plots of the accumulated phase noise for different regions of the $N-c$ landscape are presented in Figs 6, 7, and  8. Numerical simulations of uncorrelated phase noise in the region $0<N<50$ and $0<c<10$ are presented in Fig. 6. Figures 6 (a), (b), (c) correspond to phase noise with a spread $\Delta \alpha = 2\%$ for $W=1,2,3$, respectively.  Figures 6 (d), (e), (f) correspond to phase noise with a spread $\Delta \alpha = 5\%$ for $W=1,2,3$, respectively.  Figures 6 (g), (h), (i) correspond to phase noise with a spread $\Delta \alpha = 10\%$ for $W=1,2,3$, respectively. Clearly, for $\Delta \alpha > 2 \%$ fluctuations in the phase render the measurement protocol significantly unstable, and the topological quantization of the phase between $0$ and $\pi$ is blurred.
 
 Next, numerical simulations of uncorrelated phase noise in the region $100<N<500$ and $0<c<10$ are presented in Fig. 7. Figures 7 (a), (b), (c) correspond to phase noise with a spread $\Delta \alpha = 2\%$ for $W=1,2,3$, respectively.  Figures 7 (d), (e), (f) correspond to phase noise with a spread $\Delta \alpha = 5\%$ for $W=1,2,3$, respectively.  Figures 7 (g), (h), (i) correspond to phase noise with a spread $\Delta \alpha = 10\%$ for $W=1,2,3$, respectively. Clearly, this region of the $N-c$ landscape is significantly robust to uncorrelated phase noise, and the (topological) quantization of the phase is preserved. 
  
  Finally, numerical simulations of uncorrelated phase noise in the region $100<N<500$ and $100<c<400$ are presented in Fig. 8. Figures 8 (a), (b), (c) correspond to phase noise with a spread $\Delta \alpha = 2\%$ for $W=1,2,3$, respectively.  Figures 8 (d), (e), (f) correspond to phase noise with a spread $\Delta \alpha = 5\%$ for $W=1,2,3$, respectively.  Figures 8 (g), (h), (i) correspond to phase noise with a spread $\Delta \alpha = 10\%$ for $W=1,2,3$, respectively. Clearly, in this limit the only region of the $N-c$ landscape robust to phase noise if for $N>>c$ and $c<300$, we note that there is no clear critical measurement-strength parameter $c$ in this region where $\mathcal{X}$ makes a discrete jump, within the  numerical resolution required in order to cover such a large portion of the $N-c$ landscape. Numerical results considering phase noise confirm the findings in Fig. 5, with the additional insight that the $N-c$ landscape region $0 < N< 50$ and $0 < c< 10$ is not sufficiently robust to phase noise $\Delta \alpha > 2 \%$. Thus the only entirely acceptable region, regarding uncorrelated phase noise, is for $N>>c$ and $c<300$, such as the case considered in Fig. 7, and in PNAS2020 \cite{2}.

\section{Discussion}

We presented an analytical and numerical study of a novel type of geometric phase induced by weak measurements. In contrast to the case considered by Gebhart \emph{et al.} \cite{2}, we consider the dependence of the geometric phase on the winding $W$ of the
 polar angle $\varphi$, which quantifies the number of full $2\pi$-turns the trajectory makes until it closes its path, enabling to define a geometric phase $\mathcal{X}$. This is accomplished by fixing the azymhutmal angle at $\theta=\pi/2$ for a sequence of weak measurements in the angle $\varphi$ of increasing magnitude $\varphi_{k}= \epsilon k$, with $k=1,..,N$ the measurement index. 
 We ensure the trajectory induced by the sequence of weak measurements is closed, and the geometric phase well defined, by parameterizing the rotation parameter as $\alpha=2 \epsilon/N$, 
 where $N$ is the number of measurements. We find that different winding numbers $W$ can give rise to unpredictable critical measurement-strength parameters, 
 where the geometric phase becomes stochastic and the system undergoes a topological phase transition, thus confirming that the topological phase transitions for different $W$ are not topologically equivalent.  
 Furthermore, adding to the novelty of our work, we not only analyze the weak-measurement induced geometric phase by a full analytic derivation based on the exponential approximation proposed in \cite{2}, valid for $N \rightarrow \infty$, 
 but also we analyze the induced geometric phase numerically, thus enabling us to understand the finite-$N$ interplay of $\mathcal{X}$ with the measurement-strength parameter $c$, and its stability to fluctuations in the protocol. 
 
 In particular, we analyzed the impact of uncorrelated phase noise on the quantization of the geometric phase. We find that while the parameter region $N >> c$ of the $N-c$ landscape, characterized by parameter values $100 < N < 500$ and $0 < c < 10$, is significantly robust 
 to phase noise up to $\Delta \alpha \leq 10 \%$, the parameter region $0 < N < 50$ and $0 < c < 10$ can only support phase fluctuations within $\Delta \alpha < 2 \%$. For $c$ of the same order as $N$ ($N \approx c$), or $c > 300$, we find no 
 critical measurement-strength parameter where the geometric phase becomes stochastic, within the resolution of the numerical simulations. Other models for correlated noise, for instance phase noise increasing with the number of measurements $N$ and/or with the 
 measurement-strength parameter $c$ will be considered in a forthcoming work.
 
 We argue that the exhibited sharp transition in the acquired geometric phase indicates that our protocol could find relevant applications in measurement-induced manipulation and control of quantum states, for instance via the implementation of polarization switches or 
C-NOT gates, among other potential methods for quantum information processing. More specific, by tuning the coupling parameter $\eta$, for instance by increasing or decreasing the integration time $T$, which determines by the level of coupling between the detection qubit and the system qubit, it is possible to trigger a sharp transition in the geometric phase, which can be used as a quantum switch to manipulate and control quantum systems, with high precision.  Our findings also have repercussions on the understanding of  the foundations of quantum mechanics and quantum  measurement theory itself \cite{20}.

\section{Acknowledgements}

The author is grateful to Yuval Gefen, Kyrylo Snizhko and Alessandro Romito for many insightful discussions, and significant assistance in the development of the numerical codes. G. P. acknowledges financial support from PICT2015-0710  grant and Raices Programme.

\end{document}